\documentclass[aps,twocolumn,showpacs,preprintnumbers,nofootinbib,prd,superscriptaddress,groupedaddress,10pt]{revtex4-1}

\makeatletter
\def\l@subsubsection#1#2{}
\def\l@subsubsubsection#1#2{}
\makeatother

\usepackage{graphicx,amssymb,amsmath,amsthm,amsfonts,epsfig,epsf}
\usepackage[usenames]{color}
\usepackage{epstopdf}
\definecolor{darkred}{rgb}{0.5,0,0}

\usepackage{array}
\usepackage{afterpage}
\usepackage{bm}
\usepackage{dcolumn}
\usepackage[latin1]{inputenc}
\usepackage{latexsym}
\usepackage{rotating}
\usepackage{longtable}

\usepackage{enumerate}
\usepackage{tensor,multirow}
\usepackage{url}
\usepackage[linktocpage]{hyperref}
\usepackage{float}
\usepackage{graphicx}

\def\be{\begin{equation}}
\def\ee{\end{equation}}
\newcommand{\beq}{\begin{eqnarray}}
\newcommand{\eeq}{\end{eqnarray}}

\def\ba{\begin{align}}
\def\ea{\end{align}}

\newcounter{mnotecount}[section]

\renewcommand{\themnotecount}{\thesection.\arabic{mnotecount}}

\newcommand{\mnote}[1]
{\protect{\stepcounter{mnotecount}}$^{\mbox{\footnotesize
$
\bullet$\themnotecount}}$ \marginpar{
\raggedright\tiny\em
$\!\!\!\!\!\!\,\bullet$\themnotecount: #1} }

\begin{document}

\title{Quasinormal modes and Strong Cosmic Censorship}

\author{
Vitor Cardoso$^{1,2}$,
Jo\~ao L. Costa$^{3,4}$,
Kyriakos Destounis$^{1}$,
Peter Hintz$^{5}$,
Aron Jansen$^{6}$
}
\affiliation{${^1}$ CENTRA, Departamento de F\'{\i}sica, Instituto Superior T\'ecnico -- IST, Universidade de Lisboa -- UL,
Avenida Rovisco Pais 1, 1049 Lisboa, Portugal}
\affiliation{${^2}$ Perimeter Institute for Theoretical Physics, 31 Caroline Street North Waterloo, Ontario N2L 2Y5, Canada}
\affiliation{$^{3}$ Departamento de Matem\'atica, ISCTE - Instituto Universit\'ario de Lisboa, Portugal}
\affiliation{$^{4}$ Center for Mathematical Analysis, Geometry and Dynamical Systems, Instituto Superior T\'ecnico -- IST, Universidade de Lisboa -- UL,
Avenida Rovisco Pais 1, 1049 Lisboa, Portugal}
\affiliation{$^5$ Department of Mathematics, University of California, Berkeley, CA 94720-3840, USA}
\affiliation{$^6$ Institute for Theoretical Physics and Center for Extreme Matter and Emergent Phenomena,
Utrecht University, 3508 TD Utrecht, The Netherlands}
\begin{abstract}
The fate of Cauchy horizons, such as those found inside charged black holes, is  intrinsically connected to the decay of small perturbations
exterior to the event horizon. As such, the validity of the strong cosmic censorship (SCC) conjecture is tied to how effectively the exterior damps fluctuations.
Here, we study massless scalar fields in the exterior of Reissner--Nordstr\"om--de~Sitter black holes. Their decay rates are governed by quasinormal modes of the black hole. We identify {\it three} families of modes in these spacetimes: one directly linked to the photon sphere, well described by standard WKB-type tools; another family whose existence and timescale is closely related to the de Sitter horizon. Finally, a third family which dominates for near-extremally-charged black holes and which is also present in asymptotically flat spacetimes.
The last two families of modes seem to have gone unnoticed in the literature.
We give a detailed description of linear scalar perturbations of such black holes, and conjecture that SCC is violated in the near extremal regime.
\end{abstract}

\maketitle
%

\noindent{\bf{\em I. Introduction.}}
The study of the decay of small perturbations has a long history in General Relativity (GR).
An increasingly precise knowledge of the quantitative form of the decay of fluctuations is required to advance our understanding of gravitation, from the interpretation of gravitational wave data to the study of fundamental questions like the deterministic character of GR.

The well-known appearance of Cauchy horizons in astrophysically relevant solutions of Einstein's equations signals a potential breakdown
of determinism within GR---the future history of any observer that crosses such a horizon cannot be determined using the Einstein field equations and the initial data!
Nonetheless, in the context of black hole (BH) spacetimes, one expects that perturbations of the exterior region might be infinitely amplified by a blueshift mechanism,
turning a Cauchy horizon in the BH interior into a singularity/terminal boundary beyond which the field equations cease to make sense.
Penrose's Strong Cosmic Censorship (SCC) conjecture substantiates this expectation.

On the other hand, astrophysical BHs are expected to be stable due to perturbation damping mechanisms acting in the exterior region. Therefore, whether or not SCC holds true hinges to a large extent on a delicate competition between the decay of perturbations in the exterior region and their (blueshift) amplification in the BH interior.
For concreteness, let $\Phi$ be a linear scalar perturbation  (i.e., a solution of the wave equation) on a fixed
subextremal Reissner--Nordstr\"om (RN), asymptotically flat or de~Sitter (dS) BH, with cosmological constant $\Lambda\geq 0$.
Regardless of the sign of $\Lambda$, in standard coordinates, the blueshift effect leads to an exponential divergence governed by the surface gravity of the Cauchy horizon $\kappa_-$.

Now the decay of perturbations depends crucially on the sign of $\Lambda$.
For $\Lambda=0$, $\Phi$ satisfies an inverse power law decay~\cite{Price:1971fb,Dafermos:2014cua,Angelopoulos:2016wcv}
which is expected to be sufficient to stabilize the BH while weak enough to be outweighed by the blueshift amplification. Various results~\cite{Poisson:1990eh,Dafermos:2003wr,Dafermos:2012np,LukOhStrongI,LukOhStrongII} then suggest that, in this case, the Cauchy horizon will become, upon perturbation, a {\em mass inflation} singularity, strong enough
to impose the breakdown of the field equations.

For $\Lambda>0$, the situation changes dramatically.
In fact, it has been shown rigorously that, for some $\Phi_0\in\mathbb{C}$~\cite{SaBarretoZworski,BonyHaefner,Dyatlov:2011jd,Dyatlov:2013hba},
\begin{equation}
\label{waveProfile}
|\Phi-\Phi_0|\leq C e^{-\alpha t}\,,
\end{equation}
with $\alpha$  the {\em spectral gap}, i.e., the size of the quasinormal mode (QNM)-free strip below the real axis.
Moreover, this result also holds for non-linear coupled gravitational and electromagnetic perturbations of Kerr--Newman--dS (with small angular momentum)~\cite{Hintz:2016gwb,Hintz:2016KNdS}.
This is alarming as the exponential decay of perturbations might now be enough to counterbalance the blueshift amplification.
As a consequence the fate of the Cauchy horizon now depends on the relation between $\alpha$ and $\kappa_-$.
Will it still, upon perturbation, become a ``strong enough'' singularity in order to uphold SCC?

A convenient way to measure the strength of such a (Cauchy horizon) singularity is in terms of the regularity of the spacetime metric extensions it allows~\cite{Ori:2000fi,CGNS3,Earman}. For instance, mass inflation is related to inextendibility in (the Sobolev space) $H^1$ which turns out to be enough to guarantee the non-existence of extensions as (weak) solutions of the Einstein equations~\cite{Christodoulou:2008nj}, i.e., the complete breakdown of the field equations.

As a proxy for extendibility of the metric itself, we will focus on the extendibility of a linear scalar perturbation.
On a fixed RNdS, the results in~\cite{Hintz:2015jkj} (compare with~\cite{CostaFranzen}) show that $\Phi$ extends to the Cauchy horizon with regularity at least
\begin{equation}
\label{hintzRegEstimate}
H^{1/2+\beta} \,, \quad\quad\quad \beta \equiv \alpha/\kappa_-\;.
\end{equation}
Now the non-linear analysis of~\cite{Hintz:2016gwb,Hintz:2016KNdS,CGNS4,Dafermos:2017dbw} suggests that the metric will have similar
extendibility properties as the scalar field. It is then tempting to conjecture, as was done before in 
Refs.~\cite{Maeda:1999sv,Dafermos:2012np,Costa:2014yha}:
\emph{if there exists a parameter range for which $\beta>1/2$
then the corresponding (cosmological) BH spacetimes should be extendible beyond the Cauchy horizon with metric in $H^1$.} Even more strikingly, one may be able to realize some of the previous \emph{extensions as weak solutions of the Einstein equations}. This would correspond to a severe failure of SCC, in the presence of a positive cosmological constant!\footnote{The construction of bounded Hawking mass solutions of the Einstein-Maxwell-scalar field system with a cosmological constant allowing for $H^1$ extensions beyond the Cauchy horizon was carried out in~\cite{CGNS4}; these results use the stronger requirement $\beta>7/9$, but we expect $\beta>1/2$ to be sharp.}

It is also important to note that if $\beta$ is allowed to exceed unity then (by Sobolev embedding)  the scalar field extends in $C^1$; the coupling to gravity should then lead to the existence of solutions with bounded Ricci curvature. Moreover, for spherically symmetric self gravitating scalar fields, the control of both the Hawking mass and the gradient of the field is enough to control the Kretschmann scalar~\cite{CGNS3}.
We will henceforth relate $\beta<1$ to the blow up of curvature components.

At this moment, to understand the severeness of the consequences of the previous discussion,  what we are most lacking is an understanding of how the decay rate of perturbations $\alpha$ is related to $\kappa_-$. Since $\alpha$ is the spectral gap, this can be achieved by the computation of the QNMs of RNdS BHs.
The purpose of this work is to perform a comprehensive study of such modes and to discuss possible implications for SCC by determining $\beta$ throughout the parameter space of RNdS spacetimes.

\noindent{\bf{\em II. Setting.}}
We focus on charged BHs in de~Sitter spacetimes, the RNdS solutions.
In Schwarzschild-like coordinates, the metric reads
\begin{equation}
\label{RNdS_space}
ds^2=-F(r)dt^2+\frac{dr^2}{F(r)}+r^2(d\theta^2+\sin^2\theta d\phi^2)\,,
\end{equation}
where $F(r)=1-{2M}{r^{-1}}+{Q^2}{r^{-2}}-\Lambda r^2/3$.
 $M,\,Q$ are the BH mass and charge and $\Lambda$ is the cosmological constant.
The surface gravity of each horizon is then
\begin{equation}
\label{surfGrav}
\kappa_*= \frac{1}{2}|F'(r_*)|\;\;,\; *\in\{-,+,c\}\;,
\end{equation}
where $r_-<\,r_+<\, r_c$ are the Cauchy horizon, event horizon and cosmological horizon radius.

A minimally coupled scalar field on a RNdS background with harmonic time dependence can be expanded in terms of spherical harmonics,
\begin{equation}
\sum_{lm}\frac{\Phi_{l m}(r)}{r}Y_{lm}(\theta,\phi)e^{-i\omega t}\,.
\end{equation}
Dropping the subscripts on the radial functions, they satisfy the equation
\begin{equation}
\label{master_eq_RNdS}
\frac{d^2 \Phi}{d r_*^2}+\left(\omega^2-V_l(r)\right)\Phi=0\,,
\end{equation}
where we introduced the tortoise coordinate $dr_*=\frac{dr}{F}$.
The effective potential for scalar perturbations is
\begin{equation}
\label{RNdS_general potential}
V_l(r)=F(r)\left(\frac{l(l+1)}{r^2}+\frac{F^\prime(r)}{r}\right),
\end{equation}
where $l$ is an angular number, corresponding to the eigenvalue of the spherical harmonics.

We will be mostly interested in the characteristic frequencies of this spacetime, obtained by imposing the boundary conditions
\begin{equation}
 \Phi(r\to r_+)\sim e^{- i\omega r_*}\;\;,\;\;\Phi(r \to r_c)\sim e^{ i\omega r_*}\,,\label{bcs}
\end{equation}
which select a discrete set of frequencies $\omega_{ln}$, called the QN frequencies~\cite{Berti:2009kk}.
They are characterized, for each $l$, by an integer $n\geq 0$
labeling the mode number. The fundamental mode $n=0$ corresponds, by definition, to the longest-lived mode, i.e., to the frequency with the smallest (in absolute value) imaginary part.

To determine the spectral gap $\alpha$, and hence the decay rate of perturbations, we will focus on the set of all modes $\omega_{ln}$~\footnote{For $l=0$ there is a zero mode, corresponding to $\Phi_0$ in Eq.~\eqref{waveProfile}, which we ignore here.}
and set
\be
\alpha\equiv {\rm inf}_{ln}\left\{-\operatorname{Im}(\omega_{ln})\right\},\, \qquad \beta\equiv \alpha/\kappa_-  \,.\label{betaldef}
\ee
We will henceforth drop the ``$ln$'' subscripts to avoid cluttering.
In previous works, we have used a variety of methods to compute the QNMs~\cite{Berti:2009kk,GRIT}.
The results shown here were obtained mostly with the Mathematica package of~\cite{Jansen:2017oag} (based on methods developed in~\cite{Dias:2010eu}), and checked in various cases with
a variety of other methods~\cite{Berti:2009kk,GRIT,KaiLin1,Iyer:1986np}.
\begin{figure}[ht]
\includegraphics[width=0.5\textwidth]{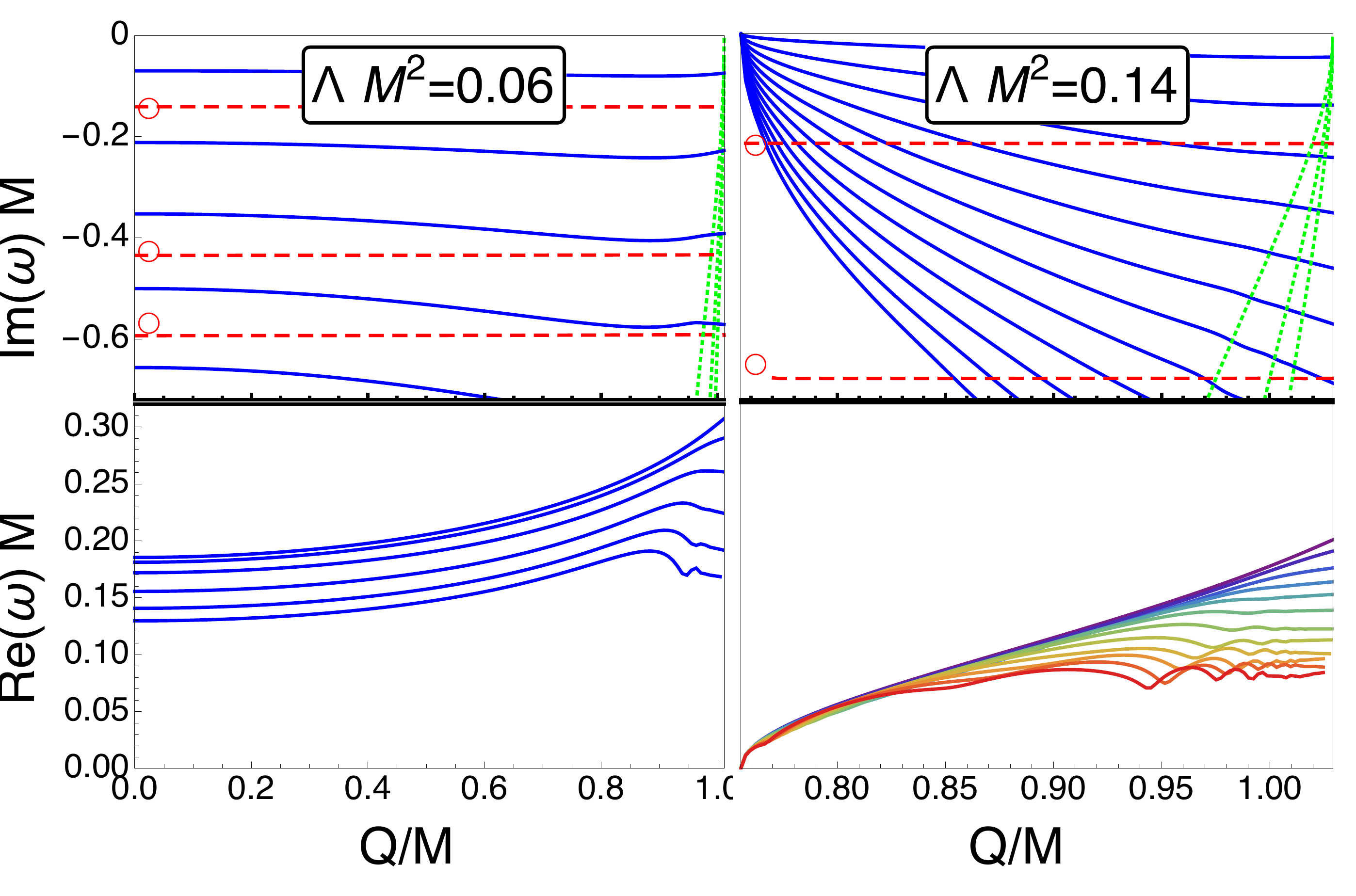}
\caption{
Lowest lying quasinormal modes for $l=1$ and $\Lambda M^2 = 0.06$ (left) and $0.14$ (right), as a function of $Q/M$.
The top plots show the imaginary part, with dashed red lines corresponding to purely imaginary modes, and solid blue to complex, ``PS'' modes, whose real part is shown in the lower plots.
The red circles in the top plots indicate the modes of empty de Sitter at the same $\Lambda$, which closely matches the first imaginary mode here, but lie increasingly less close to the higher modes.
Near the extremal limit of maximal charge, another set of purely imaginary modes (dotted green lines) comes in from $-\infty$ and approaches $0$ in the limit.
Only a finite number of modes are shown, even though we expect infinitely many complex and extremal modes in the range shown.
\label{Qdependence}}
\end{figure}

\begin{figure}[t]
\includegraphics[width=0.48\textwidth]{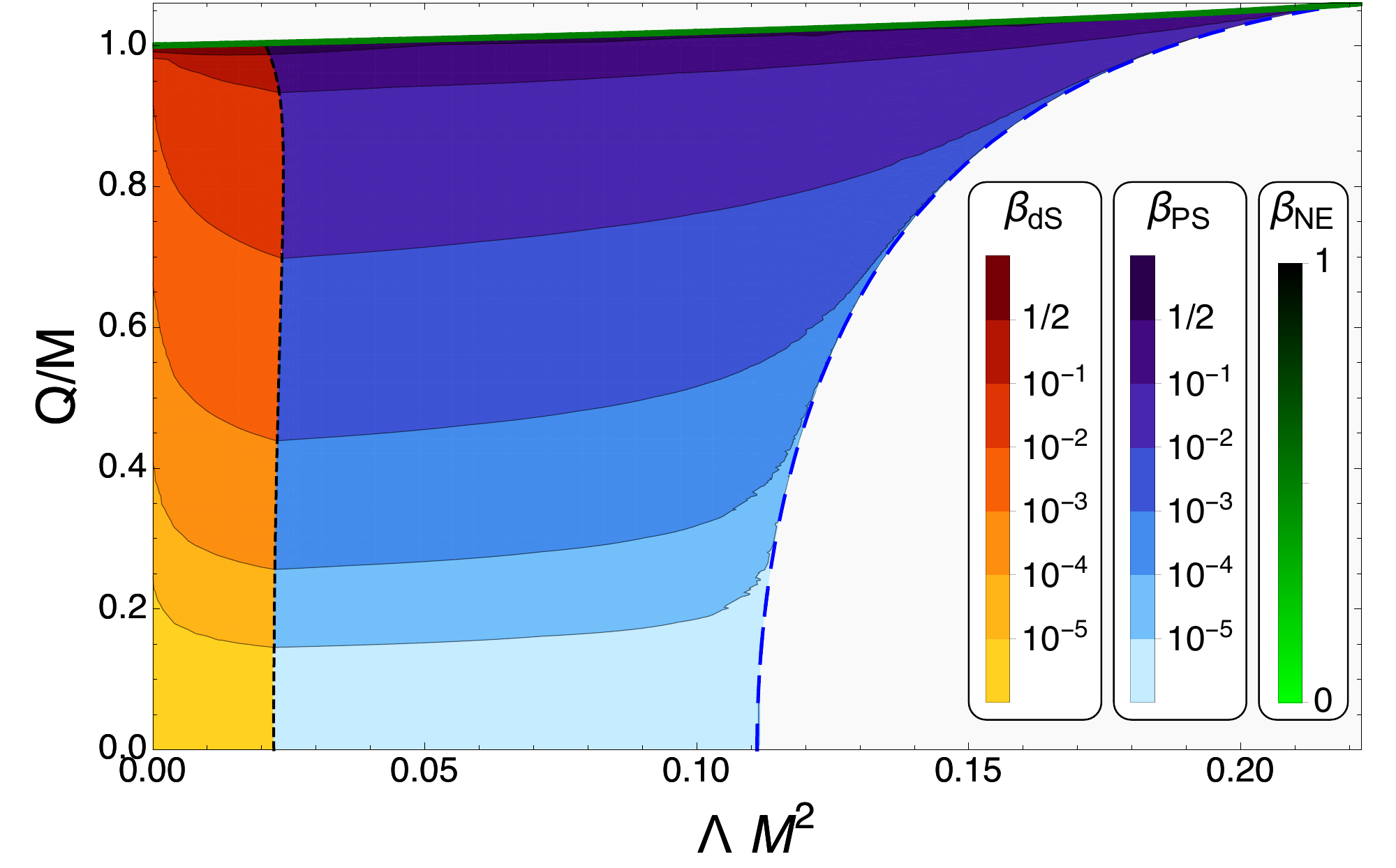}
\caption{Parameter space of the RNdS solutions, bounded by a line of extremal solutions of maximal charge where $r_- = r_+$ on top, and for $\Lambda M^2 > 1/9$ a line of extremal solutions where $r_c = r_+$.
In the physical region the value of $\beta$ is shown. For small $\Lambda M^2$ the dominant mode is the $l=1$ de Sitter mode, shown in shades of red.
For larger $\Lambda M^2$ the dominant mode is the large $l$ complex, PS mode, here showing the $l=100$ WKB approximated mode in shades of blue.
For very large $Q/M$ the $l=0$ extremal mode dominates. The green sliver on top where the NE mode dominates is merely indicative, the true numerical region is too small to be noticeable on these scales.
}
\label{ContourPlot}
\end{figure}

\begin{figure}[ht]
\includegraphics[width=0.5\textwidth]{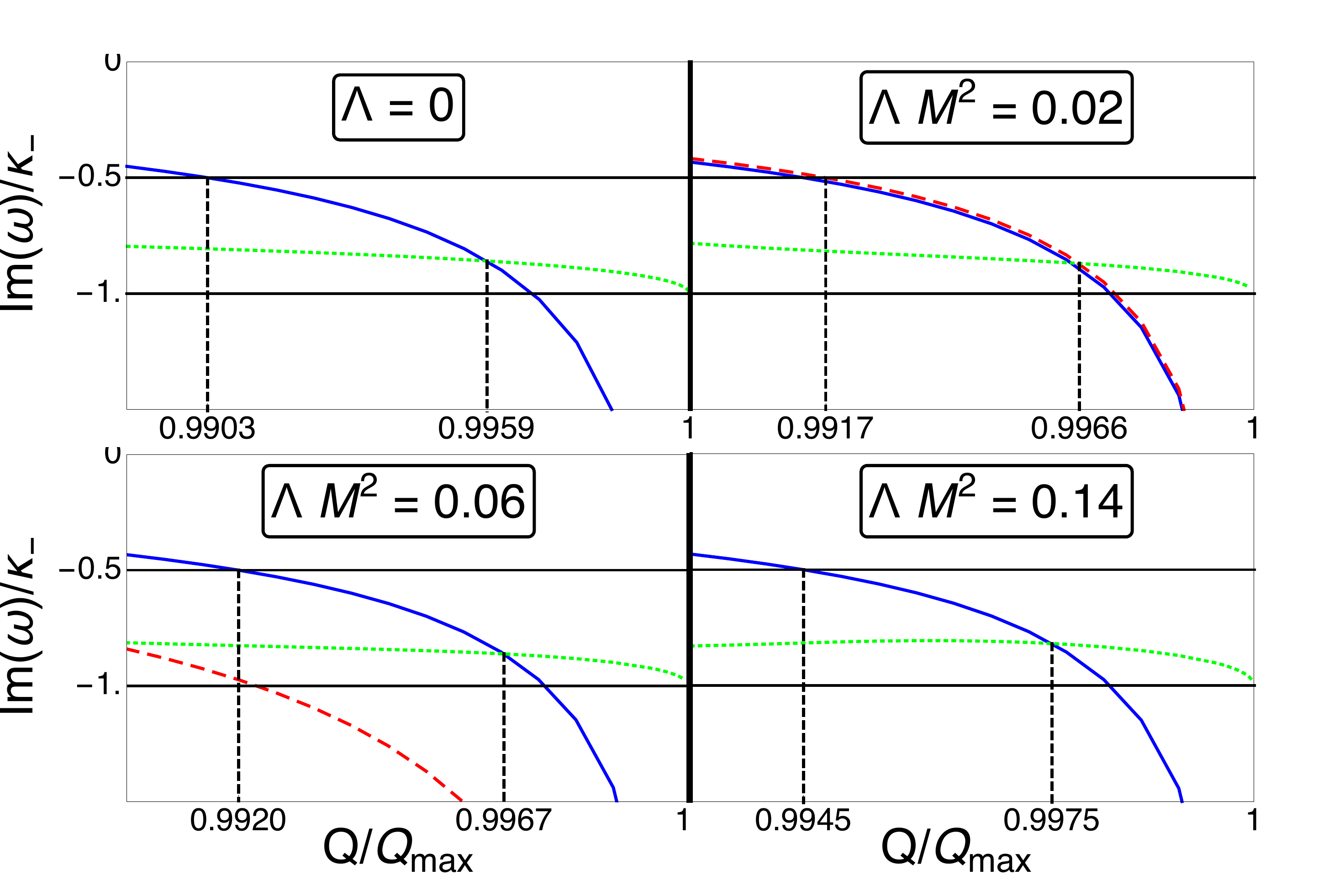}
\caption{
Dominant modes of different types, showing the (nearly) dominant complex PS mode (blue, solid) at $l= 10$, the dominant de Sitter mode (red, dotted) at $l=1$ and the dominant NE mode (green, dashed) at $l=0$.
The two dashed vertical lines indicate the points where $\beta \equiv -\operatorname{Im}(\omega)/\kappa_- =1/2$ and where the NE becomes dominant. (Note that the value of $\beta$ is only relevant for $\Lambda>0$.)
}
\label{nearExtremalModes}
\end{figure}

\noindent{\bf{\em III. QNMs of RNdS BHs: the three families.}}
Our results are summarized in Figs.~\ref{Qdependence}-\ref{nearExtremalModes} where one can distinguish three families of modes:

\noindent{\bf{\em Photon sphere modes.}}
Black holes and other sufficiently compact objects have trapping regions. Here, null particles can be trapped on circular unstable trajectories, defining the photon sphere.
This region has a strong pull in the control of the decay of fluctuations and the spacetime's QNMs which have large frequency (i.e., large $|{\rm Re}\,\omega|$)~\cite{Cardoso:2017njb,Cardoso:2008bp,PhysRevD.88.084037,Dyatlov:2011jd}. For instance, the decay timescale
is related to the instability timescale of null geodesics near the photon sphere.
For BHs in de~Sitter space, we do find a family of modes which can be traced back to the photon sphere.
We refer to them as ``photon sphere modes,'' or in short ``PS'' modes.
These modes are depicted in blue (solid line) in Figs.~\ref{Qdependence}-\ref{nearExtremalModes}. Different lines correspond to different overtones $n$; the fundamental mode
is determined by the large $l$ limit (and $n=0$); we find that $l=10$ or $l=100$ provide good approximations of the imaginary parts of the dominating mode; note however that the real parts do not converge when $l\rightarrow\infty$. These modes are well-described by a WKB approximation, and for very small cosmological constant they asymptote to the Schwarzschild BH QNMs~\cite{GRIT}.

For small values of the cosmological constant, PS modes are only weakly dependent on the BH charge. This is apparent from Fig.~\ref{Qdependence}.

For $\Lambda M^2 > 1/9$ there is now a nonzero minimal charge, at which $r_+ = r_c$. This limit is the charged Nariai BH and is shown as the blue dashed line in Fig.~\ref{ContourPlot}.
The corresponding QNMs are also qualitatively different, as seen in Fig.~\ref{Qdependence}. They in fact vanish in this limit, a result
that can be established by solving the wave equation analytically to obtain (see Ref.~\cite{Cardoso:2003sw} for the neutral case, we have generalized it to charged BHs, see Supplementary Material)
\begin{equation}
\frac{\text{Im}(\omega)}{\kappa_+} = - i \left(n + \frac{1}{2}\right) \, . \label{nariai}
\end{equation}

Note that the results presented here are enough to disprove a conjecture~\cite{BradyMossMyersStrong} that suggested that $\alpha$ should be equal to $\min\{\kappa_+,\kappa_c\}$.
Such possibility is inconsistent with~\eqref{nariai} and
it is also straightforward to find other non-extremal parameters for which the WKB prediction yields {\it smaller} $\alpha$'s  (e.g. for $\Lambda M^2=0.1$ and $Q=0$ we have $\kappa_+=0.06759$, $\kappa_c=0.05249$, and $\alpha=0.03043$).

\noindent{\bf{\em dS modes.}}
Note that solutions with purely imaginary $\omega$ exist in pure dS spacetime~\cite{LopezOrtega:2012vi,VasydS}
\begin{eqnarray}
\omega_{0, \rm pure \,\rm dS}/\kappa_c^{\rm dS} &=&-i l\,,\\
\omega_{n\neq0, \rm pure \,\rm dS}/\kappa_c^{\rm dS} &=&-i(l+n+1)\,.\label{pure_dS_scalar0}
\end{eqnarray}
Our second family of modes, the (BH) dS modes, are deformations of the pure de Sitter modes~\eqref{pure_dS_scalar0}; the dominant mode ($l=1, n=0$) is almost identical and higher modes have increasingly larger deformations.

These modes are intriguing, in that they have a surprisingly weak dependence on the BH charge and seem to be described by the surface gravity $\kappa_c^{\rm dS}=\sqrt{{\Lambda}/{3}}$ of the cosmological horizon of pure de Sitter, as opposed to that of the cosmological horizon in the RNdS BH under consideration. This can, in principle, be explained by the fact that the accelerated expansion of the RNdS spacetimes is also governed by $\kappa_c^{\rm dS}$~\cite{Brill:1993tw,Rendall:2003ks}.

This family has been seen in time-evolutions~\cite{Brady:1999wd} but, to the best of our knowledge, was only recently identified in the QNM calculation of neutral BH spacetimes~\cite{Jansen:2017oag}. Furthermore, our results indicate that as the BH ``disappears'' ($\Lambda M^2\to 0$), these modes converge to the exact de~Sitter modes (both the eigenvalue and the eigenfunction itself).

\noindent{\bf{\em Near-Extremal modes.}}
Finally, in the limit that the Cauchy and event horizon radius approach each other, a third ``near extremal'' family, labeled as $\omega_{\rm NE}$, dominates the dynamics.
In the extremal limit this family approaches
\be
\label{NEanalytic}
\omega_{\rm NE} = - i (l + n + 1) \kappa_- = - i (l + n + 1) \kappa_+ \, ,
\ee
independently of $\Lambda$, as shown by our numerics. As indicated by~\eqref{NEanalytic}, the dominant mode in this family is that for $l = 0$, this remains true away from extremality.

In the asymptotically flat case, such modes seem to have been described analytically in Refs.~\cite{Kim:2012mh,Zimmerman:2015trm,Hod:2017gvn}. Here we have shown numerically that such modes exist, and that they are in fact the limit of a new family of modes.
It is unclear (but see Ref.~\cite{Zimmerman:2015trm}) if the NE family is a charged version of the Zero-Damping-Modes discussed recently in the context of rotating Kerr BHs~\cite{Richartz:2017qep}. It is also unclear if there is any relation between such long lived modes and the instability of exactly extremal geometries~\cite{Aretakis:2012ei,Casals:2016mel}.

\noindent{\bf{\em IV. Maximizing $\beta$.}}
%
The dominating modes of the previous three families determine $\beta$, shown in Fig.~\ref{ContourPlot}.
Each family has a region in parameter space where it dominates over the other families.
The dS family is dominant for ``small'' BHs (when $\Lambda M^2\ \lesssim 0.02$).
In the opposite regime the PS modes are dominant.
Notice that in the limit of minimal charge $\beta = 0$, since $\kappa_-$ remains finite while the imaginary parts of QNMs in the PS family approach 0 according to~\eqref{nariai} (since $\kappa_+\rightarrow 0$).

More interesting is the other extremal limit, of maximal charge.
In Fig.~\ref{ContourPlot}, the uppermost contours of the dS and PS families show a region where $\beta > 1/2$.

Within this region as the charge is increased even further, the NE family becomes dominant.
In Fig.~\ref{ContourPlot} this is shown merely schematically, as the region is too small to plot on this scale, but it can already be seen in Fig.~\ref{Qdependence}.

To see more clearly how $\beta$ behaves in the extremal limit we show 4 more constant $\Lambda M^2$ slices in Fig.~\ref{nearExtremalModes}.
Here one sees clearly how above some value of the charge $\beta>1/2$, as dictated by either the de Sitter or the
PS family. Increasing the charge further, $\beta$ would actually diverge if it were up to these two families ($\omega M$ approaches a constant for both families, so $\omega/\kappa_-$ diverges).
However, the NE family takes over to prevent $\beta$ from becoming larger than 1. Further details on these modes and on the maximum $\beta$ are shown in the Supplementary Material.
%

\noindent{\bf{\em V. Conclusions.}}
The results in~\cite{Hintz:2016gwb,Hintz:2016KNdS} show that the decay of small perturbations of de Sitter BHs is dictated by the spectral gap $\alpha$. At the same time, the linear analysis in~\cite{Hintz:2015jkj} and the non-linear analysis in~\cite{CGNS4} indicate that the size of $\beta\equiv\alpha/\kappa_-$ controls the stability of Cauchy horizons and consequently the fate of the SCC conjecture.  Recall that for the dynamics of the Einstein equations, and also for the destiny of observers, the blow up of curvature (related to $\beta<1$) per se is of little significance: it implies neither the breakdown of the field equations~\cite{L2} nor the destruction of macroscopic observers~\cite{Ori:1991zz}.
In fact, a formulation of SCC in those terms is condemned to overlook relevant physical phenomena like impulsive gravitational waves or the formation of shocks in relativistic fluids. For those and other reasons, the modern formulation of SCC, which we privilege here, makes the stronger request $\beta<\frac{1}{2}$ in order to guarantee the breakdown of the field equation at the Cauchy horizon.

Here, by studying (linear) massless scalar fields and searching through the entire parameter space of subextremal and extremal RNdS spacetimes, we find ranges for which $\beta$ exceeds $1/2$ but, remarkably, it doesn't seem to be allowed, by the appearance of a new class of ``near-extremal'' modes, to exceed unity! This opens the perspective of having Cauchy horizons which, upon perturbation, can be seen as singular, by the divergence of curvature invariants, but nonetheless maintain enough regularity as to allow the field equations to determine (classically), in a highly non-unique way, the evolution of gravitation. This corresponds to a severe failure of determinism in GR that cannot be taken lightly in view of the importance that a cosmological constant has in cosmology and the fact that the pathologic behavior is observed in parameter ranges which are in loose agreement with what one expects from the parameters of some astrophysical BHs~\footnote{Astrophysical BHs are expected to be neutral and here we are dealing with charged BHs. This is justified by the standard charge/angular momentum analogy, where near-extremal charge corresponds to fast rotating BHs.}~\cite{2011ApJ...736..103B,Middleton:2015osa,FastSpin}.

\noindent{\bf{\em Acknowledgments.}}
V.C. is indebted to Kinki University in Osaka and to the Kobayashi-Maskawa Institute in Nagoya for hospitality, while the late stages of this work were being completed.
V.C. and K.D. acknowledge financial support provided under the European Union's H2020 ERC Consolidator Grant ``Matter and strong-field gravity:
New frontiers in Einstein's theory'' grant agreement no. MaGRaTh--646597. Research at Perimeter Institute is supported by the Government of
Canada through Industry Canada and by the Province of Ontario through the Ministry of Economic Development $\&$
Innovation.
J.L.C. acknowledges financial support provided by FCT/Portugal through UID/MAT/04459/2013 and grant (GPSEinstein) PTDC/MAT-ANA/1275/2014.
A.J. was supported by the Netherlands Organisation for Scientific Research (NWO) under VIDI grant 680-47-518, and the Delta-Institute for Theoretical Physics (D- ITP) that is funded by the Dutch Ministry of Education, Culture and Science (OCW).
This project has received funding from the European Union's Horizon 2020 research and innovation programme under the Marie Sklodowska-Curie grant
agreement No 690904.
Part of this research was conducted during the time P.H.\ served as a Clay Research Fellow; P.H.\ also acknowledges support from the Miller Institute at UC Berkeley.
The authors would like to acknowledge networking support by the COST Action GWverse CA16104.
The authors thankfully acknowledge the computer resources, technical expertise and assistance provided by S\'ergio Almeida at CENTRA/IST. Computations
were performed at the cluster ``Baltasar-Sete-S\'ois'', and supported by the MaGRaTh--646597 ERC Consolidator Grant.

\bibliography{references}

\vskip 2cm

\clearpage
{\bf{\em \Large Supplementary Material}}

\begin{figure}[t]
\includegraphics[width=.48\textwidth]{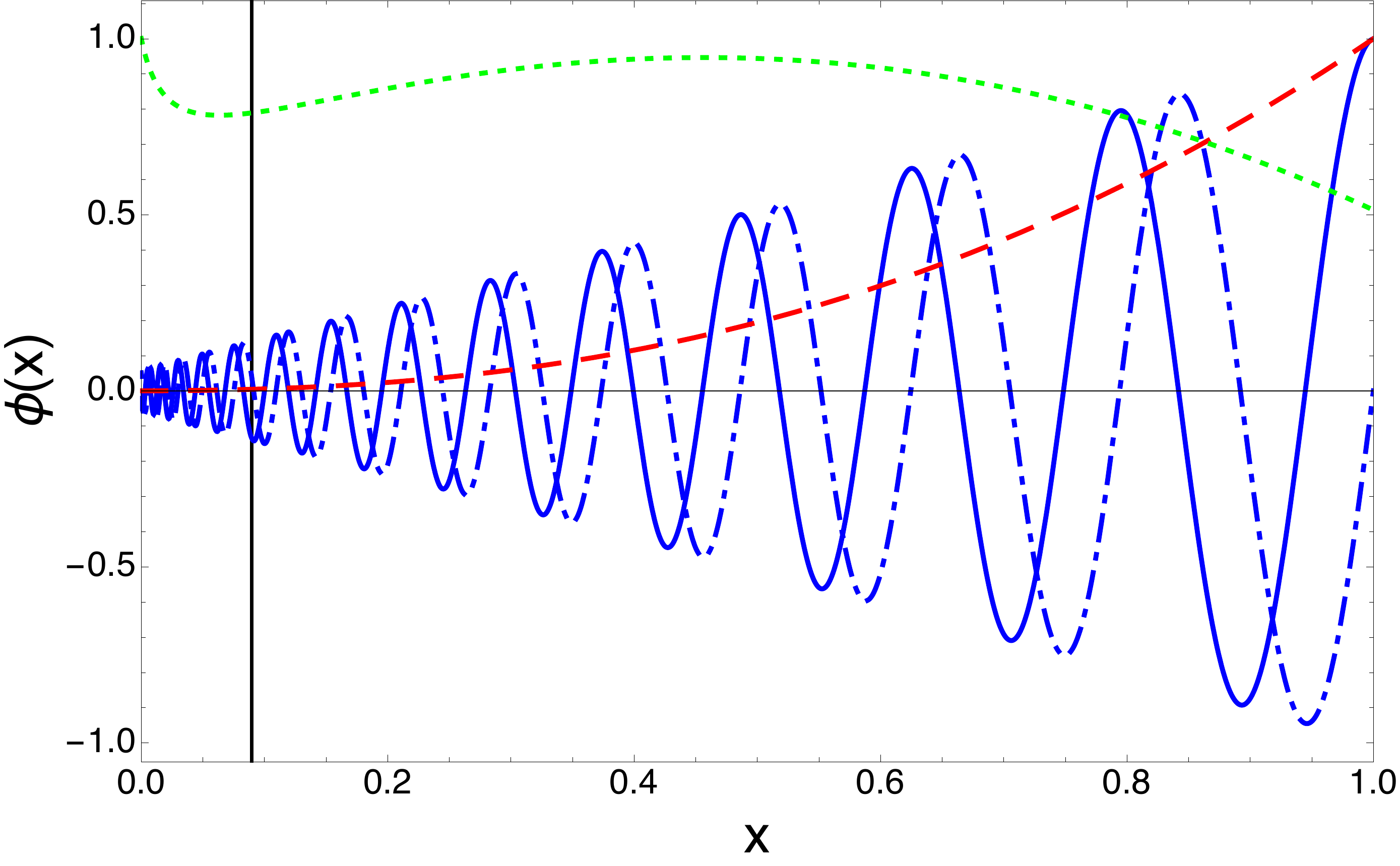}
\caption{
Scalar wavefunctions $\phi(x)$ (defined in Eq.~(\ref{phidef}), with $x = (r - r_+)/(r_c - r_+)$) of the dominant mode of each of the three families, at $\Lambda M^2 = 0.02$ and $Q/Q_{\text{max}} \approx 0.9966$, as indicated in the top right of Fig. (\ref{nearExtremalModes}).
Shown are the near extremal mode with $\omega_{\text{NE}}/\kappa_- = -0.87005i$ (for $l=0$, green dotted line), 
the dS mode with $\omega_{\text{dS}}/\kappa_- = -0.87043i$ (for $l=1$, red dashed line), 
and the complex mode with $\omega_{\text{C}}/\kappa_- = 26.448 -0.89101i$ (for $l=10$, blue solid and dash-dotted line for real and imaginary parts). The solid vertical black line indicates the light ring.
}
\label{eigenfunctionPlot}
\end{figure}

\noindent {\bf{\em The eigenfunctions}}
The difference between PS, dS and NE modes is also apparent from the eigenfunction itself. It is useful to define a re-scaled function
$\phi(r)$ as,
\begin{equation}
\label{phidef}
\Phi(r) = (r - r_+)^{- i \omega / (2 \kappa_+) }  \phi(r)  (r_c - r)^{- i \omega / (2 \kappa_c) } \,.
\end{equation}
The conditions on $\phi(r)$ are that it approaches a constant as $r \rightarrow r_+, r_c$,
Figure~\ref{eigenfunctionPlot} shows the behavior of $\phi$ for different modes, for a specific set of RNdS parameters.
Although not apparent, there is structure close to the photon sphere for the PS eigenfunction.

\noindent {\bf{\em Analytic solutions for $r_c = r_+$}}
In the limit $r_c = r_+$, the limit of minimal charge for $\Lambda M^2 \geq 1/9$, the quasinormal modes can be found analytically.
In this limit, equation (\ref{master_eq_RNdS}) with potential (\ref{RNdS_general potential}), written in the coordinate $x=(r-r_+)/(r_c-r_+)$, becomes
\begin{widetext}
\begin{equation}
\left( \frac{4  r_+ x (1-x) }{3-2 r_+} l (l+1) + \lambda^2 \right) \phi(x) + 4 x (1-x) (1-2 x) \phi^\prime(x) + 4 x^2 (1-x)^2  \phi^{\prime\prime}(x) = 0 \,
\end{equation}
\end{widetext}
where we defined $\lambda \equiv \omega / \kappa_+$ and we have set $M=1$, which can be restored in the end by dimensional analysis.

This equation has the solutions
\begin{equation}
\phi(x) = c_1 P_{\alpha }^{i \lambda }(2 x-1)+c_2 Q_{\alpha }^{i \lambda }(2 x-1) \, ,
\end{equation}
where $P, Q$ are the Legendre $P$ and $Q$ functions, and
\begin{equation}
\alpha = \frac{1}{2} \left( -1 + \sqrt{1 - \frac{2 r_+}{r_+ - 3/2} l (l+1)} \right) \, .
\end{equation}

Now, analysis of the asymptotic behavior near $x=0$ and $x=1$ shows that we can only satisfy the ``outgoing'' boundary conditions~\cite{Berti:2009kk} 
when $c_2 = 0$ and $\lambda$ is either $\lambda = - i (\alpha + n)$, or $\lambda = -i (n + 1 - \alpha)$.
These combine to give
\begin{equation}
\frac{\omega}{\kappa_+} = \pm \frac{1}{2} \sqrt{-1 +  l(l+1) \frac{2 r_+}{r_+ - 3/2}} - i \left( n + \frac{1}{2}\right) \, ,
\end{equation}

Restoring units we obtain,
\begin{eqnarray}
\label{modeMinQ}
\frac{\omega}{\kappa_+} &=& \pm \frac{1}{2} \sqrt{-1 + 2 l (l+1) \Upsilon}  - i \left(n + \frac{1}{2}\right) \, , \label{nariai} \\
\Upsilon&=&\frac{\gamma^2 + 2^{1/3} \Lambda M^2}{\gamma^2 +( 2^{1/3} - 3 \times 2^{-1/3} \gamma) \Lambda M^2}\,, \nonumber \\
\gamma & =& \left( - 3 (\Lambda M^2)^2 + (\Lambda M^2)^{3/2} \sqrt{(9 \Lambda M^2 - 2)} \right)^{1/3} \,. \nonumber
\end{eqnarray}
The argument of the square root in \eqref{nariai} is positive.
The imaginary part of these frequencies is exactly the same as that of the neutral Nariai BH~\cite{Cardoso:2003sw}.
The real part is different and now depends also on $\Lambda$, but asymptotes to the previous result for $Q=0$ (i.e., $\Lambda = 1/9$), as it should. It is an interesting feature that when $\Lambda M^2 \rightarrow 2/9$, $\Upsilon$, and therefore the real part of $\omega/\kappa_+$, diverges.

%

\noindent {\bf{\em Searching for long-lived modes}}
Here we address the question whether there might be a more slowly decaying mode that we have missed and could save SCC.
If such a mode exists, it would be highly unlikely to be part of the three families we found, since we can follow their continuous change as the BH parameters are varied, as shown in the figures.
Furthermore, the known modes in the limiting cases are all accounted for, and we never observed any mode crossings for given family and angular momentum $l$.
\begin{table}[h!]
\begin{center}
\begin{tabular}{|c||c|}
 $l$ & $\omega _ 0/\kappa _-$ \\ \hline\hline
 0 & \underline{\bf-0.8539013779 i(61)} \\ \hline
 1 & 3.2426164126 - 0.7958326323 i(67) \\ 
     & ( \underline{\bf -1.5003853731 i(64)}) \\ \hline
 2 & 5.4796067815 - 0.7754179185 i(69) \\ \hline
 3 & 7.7016152057 - 0.7699348317 i(70) \\ \hline
 4 & 9.9181960834 - 0.7676996545 i(73) \\ \hline
 5 & 12.1322536641 - 0.7665731858 i(75) \\ \hline
 10 & \underline{\bf23.1897597770 - 0.7649242108 i(80)} \\ \hline
 100 & 222.0602900249 - 0.7643094446 i(82) \\ \hline
\end{tabular}
\end{center}
\caption{
The dominant quasinormal modes for a range of angular momenta $l$, in units of the surface gravity of the Cauchy horizon, for the BH with $\Lambda M^2 = 0.06$ and $Q/Q_\text{max} = 0.996$.
The bold, underlined modes at $l=0, 1$ and $10$ are the dominant modes for the near-extremal-, de Sitter- and photon sphere modes respectively, as seen also in the bottom left of Fig. 3.
Note that the de Sitter mode is subdominant even for fixed $l=1$, and the photon sphere mode at $l=10$ is only dominant to very good approximation, the true dominant mode being that with $l \rightarrow \infty$.
Numbers in brackets indicate the number of agreed digits in the computations with grid size and precision $(N,p) = (400,200)$ and $(450,225)$.
}
\label{extracheck}
\end{table}

It is theoretically possible, though also very unlikely, that there is a fourth family (with anywhere between a single and infinitely many members) that we have missed.

Typically, smaller eigenvalues are found more easily than larger eigenvalues, making it more unlikely to miss the dominant mode.
It could be however that the corresponding eigenfunction is either very sharply peaked or highly oscillatory, in which cases it would require a large number of grid points to be resolved accurately enough.
This again decreases the possibility that we have missed something. We will rule out this last scenario, as best as we can numerically, as follows. We pick a representative BH for which $\beta > 1/2$, indicating violation of SCC, namely $\Lambda M^2 = 0.06$ and $Q/Q_\text{max} = 0.996$. For these BH parameters we compute the QNMs for various angular momenta $l$ as shown in Table~\ref{extracheck}.
There is no new QNM that is more dominant than those found before, except as expected the $l= 100$ photon sphere mode, but not significantly, note the extremely rapid convergence with increasing $l$.

The main method we use essentially discretizes the equation and rearranges it into a generalized eigenvalue equation, whose eigenvalues are the QNMs (see~\cite{Jansen:2017oag} for more details). 
This has two technical parameters, the number of grid points $N$ and the precision $p$ (number of digits) used in the computation. 
To be sure that the obtained results are not numerical artefacts one has to repeat the computation at different $(N,p)$ and test for covergence, which we have done for all results shown.

The computation here was done at even higher accuracy than in the main results, with $(N,p) = (400,200)$ and $(450,225)$.
The most we used previously was  $(300,150)$ and $(350,175)$ (near extremality, away from extremality a much lower accuracy usually suffices). The number in brackets behind each mode in Table~\ref{extracheck} is the number of digits that agrees between the computations at these two accuracies. 

We checked that even before testing for convergence, there are no modes with imaginary part smaller (in absolute sense) than shown in Table~\ref{extracheck}.
This confirms our results with as much certainty as one can reasonably expect from a numerical result.

\end{document}